\documentclass[preprint,nofootinbib,showpacs,superscriptaddress,aps,prc]{revtex4-2}
\usepackage{graphicx}
\usepackage{amstext}
\usepackage{amssymb}
\usepackage[usenames]{color}

\begin{document}
\title{Kaon and pion maximal emission times extraction \\ from the femtoscopy analysis  
of $5.02 A$~TeV LHC collisions \\ within the integrated hydrokinetic model}

\author{V.~M.~Shapoval}
\author{Yu.~M.~Sinyukov} 
\affiliation{Bogolyubov~Institute~for~Theoretical~Physics, Metrolohichna str.~14b, 03143 Kyiv, Ukraine}

\begin{abstract}
A simple method for the extraction of the times of maximal emission for kaons and pions using
the combined fitting of their transverse momentum 
spectra and the longitudinal interferometry radii dependencies on the pair transverse mass $m_T$ is applied to Pb+Pb 
collisions at the LHC energy $\sqrt{s_{NN}}=5.02$~TeV. The method is based on the analytical formulas, that were 
earlier successfully utilized in the studies of Pb+Pb collisions at $\sqrt{s_{NN}}=2.76$~TeV. 
To test the method, the spectra, radii and particle radiation picture are calculated within the integrated 
hydrokinetic model (iHKM), that includes all the stages of the matter evolution in high-energy A+A collisions: 
the system's formation, its thermalization, viscous hydrodynamics evolution, particlization and subsequent 
hadronic cascade. 
The model describes and predicts well already published LHC data in ``soft physics'' kinematic region. 
Thus, the fitting results for maximal emission times of kaons and pions are compared to the approximate maximal 
emission time values, estimated based on the emission function plots, obtained in iHKM. The developed simple method   
is intended for use in experimental analysis of femtoscopy data in relativistic A+A collisions.  
\end{abstract}

\maketitle



\section{Introduction}

The well-developed experimental technique of correlation femtoscopy measurements in the field of
high-energy heavy-ion collision physics makes it possible to investigate the spatio-temporal
structure of the systems produced in such collisions, as well as the peculiarities of the process
of their evolution (see, e.g. review~\cite{Lisa}). The femtoscopy, or interferometry radii, extracted from the Gaussian fits to the
measured two-particle correlation functions at given pair momentum~$k_T$, are generally associated with
the homogeneity lengths of a rapidly expanding system, or the three-dimensional sizes of the fragment of the system, 
where the particles with momentum~$k_T$ are mainly emitted from~\cite{hlength1,hlength2}.
The detailed structure of homogeneity lengths contains also the information on, e.g., the strength of the 
developed collective flow and the lifetime of the created fireball~\cite{MakSin} and on 
the space-time correlations between emitted particles~\cite{hbt-puzzle1,hbt-puzzle2}. 
The femtoscopy analysis can provide even such advanced information about the dynamics of the system's expansion as
the effect of the resonance decays and hadron-hadron scatterings at the 
afterburner stage of the collision on the formation of bulk observables, or the times at which the particles of 
different species are mostly emitted from the system~\cite{lifetime,sourcefunc}. The study of the correlation
functions for pairs of non-identical particles can help to find out particles of which sort are emitted
earlier~\cite{lednicky,kiesel}.

In the paper~\cite{lifetime} we proposed a method for the estimation of the times of maximal emission
for pions and kaons in the LHC Pb+Pb collisions at $2.76 A$~TeV based on a simple analytical formulas allowing the simultaneous fitting
of transverse momentum spectra for both considered particle sorts followed by fitting of the corresponding 
$long$ interferometry radii dependencies on pair $m_T$. The formulas were obtained
as a result of analytical approximation for single-particle and two-particle momentum spectra in A+A collisions \cite{hlength2,lifetime}. 
This method gave us the estimates for the effective pion and kaon emission times within the hydrokinetic model 
(HKM)~\cite{HKM,HKM1}, which was used to calculate the particle spectra in our study. 
The obtained estimates were also in agreement with the corresponding emission function plots obtained in HKM. 
In addition, the first application of the method to the RHIC BES energies ~\cite{Oslo}, where the 
UrQMD was used as the evolutionary model, was quite satisfactory. 

The method was  
successfully used by the ALICE Collaboration~\cite{alice-mt} for the estimation of pion and kaon maximal
emission times within their experimental analysis. Both studies showed that the effective emission time
for kaons is larger than for pions. This fact, together with observed essentially non-Gaussian shape of
the considered correlation functions and the absence of scaling between $R_{\mathrm{long}}(m_T)$ dependencies 
for kaons and pions, suggested that the secondary kaons, coming from the $K^{*}(892)$ resonance decays and involved
into intensive interaction with the hadronic medium at the final stage of the collision, give an important
contribution to the total kaon yield. Similar results were obtained later within the more advanced 
integrated hydrokinetic model (iHKM)~\cite{ihkm1,ihkm2} for the case of Au+Au collisions at the top RHIC energy 
in our recent paper~\cite{rhic-ihkm}.

In~\cite{lhc502-ihkm} we presented the results of our systematic study of Pb+Pb collisions at the LHC energy 
$5.02 A$~TeV within the iHKM model concerning the most of the bulk observables, including the predictions
for femtoscopy radii in the three centrality classes ($c=0-5\%$, $c=20-30\%$, and $c=40-50\%$). 
However, our previous study did not include
the analysis of the maximal emission times for kaons and pions. That is why in the present work we are going
to close this gap and to apply the developed technique to the case of Pb+Pb collisions at $\sqrt{s_{NN}}=5.02$~TeV
in order to get more details about the character of the matter evolution at this LHC energy and to compare it
with the case of $\sqrt{s_{NN}}=2.76$~TeV collisions.

\section{Analytical model}

In this section we briefly explain the idea of the method utilized (see~\cite{lifetime} for details)
and the origination of the analytical formulas for spectra and radii fitting.

The formation of particle spectra measured in heavy-ion collision experiments can be described
using a modified Cooper-Frye prescription (CFp) for particlization hypersurface~$\sigma$, which consists 
of all space-time
points $(t_{\sigma}(\textbf{r},p),\textbf{r})$, where the maximal emission 
of quanta with momentum $p$ takes place~\cite{spec-form1,spec-form2}. In this approach, by contrast
with the standard Cooper-Frye prescription, where the same particlization hypersurface (typically an isotherm) 
is used for all the momenta $p$, one does not have any problems with the negative contributions
from non-spacelike elements of the switching hypersurface, since for each specified momentum $p$ the
corresponding fragment of the hypersurface is spacelike by its construction.

Here, similarly to~\cite{lifetime}, we base our consideration on the approximation~\cite{tau-const1,tau-const2} 
of such generalized CFp, where we suppose the hypersurface $\sigma$ to be of constant proper 
time $\tau$ (equal to the time of maximal emission for soft quanta, $\tau=\tau_{m.e.}=const$) and  
to be limited in the direction $\textbf{r}_T$, transverse to the beam axis. 
Such an assumption corresponds to the emission of soft particles with momenta $p_T \approx 0.2-0.4$~GeV/$c$,
whereas for particles with $p_T>0.8$~GeV/$c$ strong space-time correlations between particle radiation events
take place, and the related hypersurface parts essentially differ from $\tau=const$~\cite{tau-const1,tau-const2}.
Thus, we utilize the following analytical representation for the bosonic Wigner function for soft enough quanta:
\begin{equation}
f_{l.eq.}(x,p)=\frac{1}{(2\pi)^3}\left[\exp(\beta p\cdot u(\tau_{m.e.},{\bf r}_T) -\beta\mu)-1\right]^{-1}\rho({\bf r}_T),
\label{wigner}
\end{equation} 
where $\beta$ denotes the inverse temperature, $\eta_L= \text{arctanh}\,v_L$
and $\eta_T = \text{arctanh}\,v_T(r_T)$ are longitudinal and transverse rapidities,
$u^{\mu}(x)=(\cosh\eta_L\cosh\eta_T,\frac{{\bf r}_T}{r_T}\sinh\eta_T,\sinh\eta_L\cosh\eta_T)$ is 
hydrodynamic velocity and $\rho({\bf r}_T)$ is the Gaussian cutoff factor
\begin{equation}
\rho({\bf r}_T)=\exp[-\alpha (\cosh\eta_T(r_T)-1)],
\label{rho}
\end{equation}
where the parameter $\alpha$ is defined as $\alpha = R_v^2/R_T^2$, 
and in the latter ratio $R_T$ denotes the homogeneity length
in transverse direction $\textbf{r}_T$ (for $r_T$ close to zero and at small momenta~$k_T$),
while $R_v=(v^{\prime}(r_T))^{-1}$ is the hydrodynamic length near $r_T=0$.

The $\alpha$ parameter characterizes the strength of the collective flow: strong flow corresponds to
small values of $\alpha$, since in this case one has $R_v<<R_T$, and for the case of absent flow 
$R_v \rightarrow \infty$, so that $\alpha \rightarrow \infty$ as well.
The factor $\rho({\bf r}_T)$ effectively limits the particlization hypersurface in transverse direction 
and removes the contributions from hard quanta, for which $\cosh\eta_T(r_T) \gg 1$.

According to the improved Cooper-Frye prescription utilized in our analysis one can
calculate the single-particle spectra $p_{0}(d^{3}N/d^{3}p)$ 
and two-particle correlation functions $C(p,q)$ as follows:
\begin{equation}
p_{0}\frac{d^{3}N}{d^{3}p}=\int_{\sigma_{m.e.}(p)}d\sigma_{\mu} p^{\mu} f_{l.eq.}(x,p),
\label{sp-def}
\end{equation}
\begin{equation}
C(p,q)\approx 1+\frac{\left|\int_{\sigma_{m.e.}(k)} d\sigma_{\mu}k^{\mu} f_{l.eq.}(x,k)\exp(iqx)\right|^{2}}{\left(\int_{\sigma_{m.e.}(k)}
d\sigma_{\mu}k^{\mu} f_{l.eq.}(x,k)\right)^{2}}.
\label{corr-approx}
\end{equation}
Here we use the conventional denotations $q=p_{1}-p_{2}$ and $k^{\mu}=\left(\sqrt{m^2+\left(\frac{\mathbf{p_{1}}+\mathbf{p_{2}}}{2}\right)^2},\frac{\mathbf{p_{1}}+\mathbf{p_{2}}}{2}\right)$.
When one works in both smoothness and mass shell approximations, one has $k\approx p=(p_1+p_2)/2$
and 4-vector $q$ having only three independent components. In femtoscopy analysis one usually chooses
them to be $q_{\mathrm{long}}$ --- along the beam axis direction, $q_{\mathrm{out}}$ --- along
the pair transverse momentum $\textbf{k}_T$ direction, and $q_{\mathrm{side}}$ --- orthogonal
to both {\it long} and {\it out} directions.

To obtain some simple analytical expressions for spectrum and correlation function,
which would be convenient for fitting and interpretation of the experimental or the realistic simulation
results, one can approximately calculate Eq.~(\ref{sp-def}) and Eq.~(\ref{corr-approx}),
substituting the Wigner function $f_{l.eq.}(x,p)$ by (\ref{wigner}) and
using the saddle point method, as it was done in~\cite{Tolstykh} within the Boltzmann approximation
for the case of longitudinally boost-invariant expansion.

It is interesting, that within the obtained in~\cite{Tolstykh} approximation the behavior of the correlation function 
$C(p,q)$ in {\it long} direction is defined only by the $\alpha$ parameter value and does not depend on the profile
of transverse velocity $v_T$ at the particlization hypersurface (unfortunately, this is not the case for the two
transverse directions). This allows to simplify further analysis significantly, if in what follows we restrict
ourselves to the analysis of longitudinal direction only.

Introducing the longitudinal homogeneity length at non-zero transverse flow, $\lambda_l=\tau\sqrt{\frac{T}{m_T}(1-\bar{v}^2_T)^{1/2}}$~\cite{tau-const1,tau-const2} and denoting its ratio to $\tau$ as $\lambda$, one has
\begin{equation}
\lambda^2 =\frac{\lambda_l^2}{\tau^2}=\frac{T}{m_T}(1-\bar{v}^2_T)^{1/2}.
\label{lambda}
\end{equation}
Here $\bar{v}_T=k_T/(m_T+\alpha T)$ is the transverse flow velocity in the saddle point, and $T=T_{m.e.}$
is the temperature at the hypersurface of maximal emission, $\tau=\tau_{m.e.}$.
Then in LCMS frame one can write for the correlation function in $l\equiv long$ direction (at $q_{\mathrm{out}}=q_{\mathrm{side}}=0$)~\cite{Tolstykh}:
\begin{equation}
C(k,q_l)=1+\frac{\exp\left[\frac{2}{\lambda^2}\left(1-\sqrt{1+\tau^2\lambda^4q_l^2}\right)\right]}{\left[1+\tau^2\lambda^4q_l^2\right]^{3/2}}
\stackrel{k_T\rightarrow \infty}{\longrightarrow} 1+\exp(-\lambda_l^2 q_l^2).
\label{correlator}
\end{equation}
The obtained expression implies that in general case the correlation function is not Gaussian,
which means that one can obtain different analytical approximations for the Gaussian interferometry radii 
based on this function, depending on the considered physical situation (see~\cite{Tolstykh} for more details).

In particular, when one considers the case of longitudinally boost-invariant matter expansion
with a transverse flow having arbitrary velocity profile $v_T(\textbf{r}_T)$, then for 
small $q_{\mathrm{long}}$ (the peak of the correlation function) one has for $R_{\mathrm{long}}(m_T)$:
\begin{equation}
R^2_{\mathrm{long}}(m_T)=\tau^2\lambda^2\left(1+\frac{3}{2}\lambda^2\right),
\label{radfit}
\end{equation}
where $m_T=\sqrt{m^2+k_T^2}$.
The formula~(\ref{radfit}) can be applied in case of transverse flow of any 
intensity, which is especially important for the LHC energies.

Having obtained the formula~(\ref{radfit}) for the femtoscopy radii fitting from the 
approximation for the correlation function (\ref{corr-approx}), one can further
use a similar approach to get the formula for momentum spectrum starting from 
the Eq.~(\ref{sp-def})~\cite{tau-const2}. As a result one will have
\begin{equation}
p_0 \frac{d^3N}{d^3p} \propto \exp{[-(m_T/T + \alpha)(1-\bar{v}^2_T)^{1/2}]}.
\label{specfit}
\end{equation}
This formula allows one to approximate the slope of the transverse momentum spectrum at not very high $p_T$
in the presence of transverse collective flow under the assumption that the shape
of the spectrum is close to the exponential one.

The procedure, proposed in~\cite{lifetime} for the estimation of the time of maximal emission
for pions and kaons in the LHC Pb+Pb collisions, suggests at first to determine the parameters $T$ and $\alpha$
based on the results of combined fitting of pion and kaon $p_T$ spectra with Eq.~(\ref{specfit}) 
and then use the found parameter values
to fit the corresponding $R_{\mathrm{long}}(m_T)$ dependencies with Eq.~(\ref{radfit}). 
This latter fitting gives one the desired time of maximal emission values $\tau_{\pi}$ and $\tau_{K}$.

\section{Results and discussion}

\begin{figure}
\centering
\includegraphics[bb=0 0 567 411, width=0.85\textwidth]{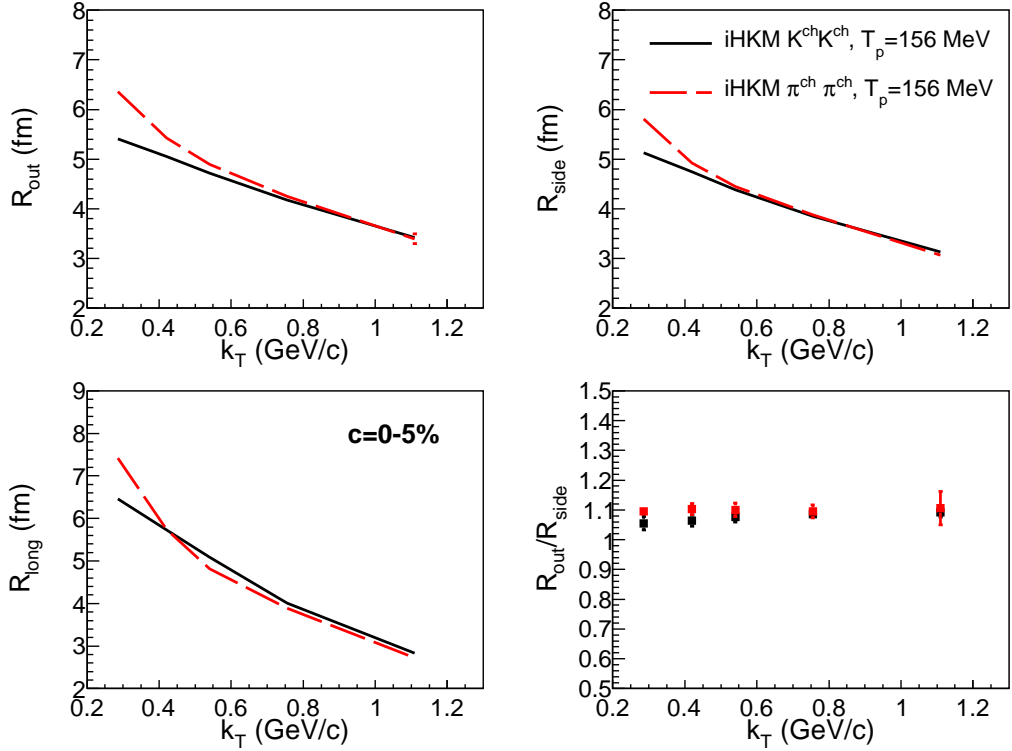}
\caption{The iHKM results for pion and kaon interferometry radii in the LHC Pb+Pb collisions
at $\sqrt{s_{NN}}=5.02$~TeV, $c=0-5\%$.
\label{rad05}} 
\end{figure}

\begin{figure}
\centering
\includegraphics[bb=0 0 567 411, width=0.85\textwidth]{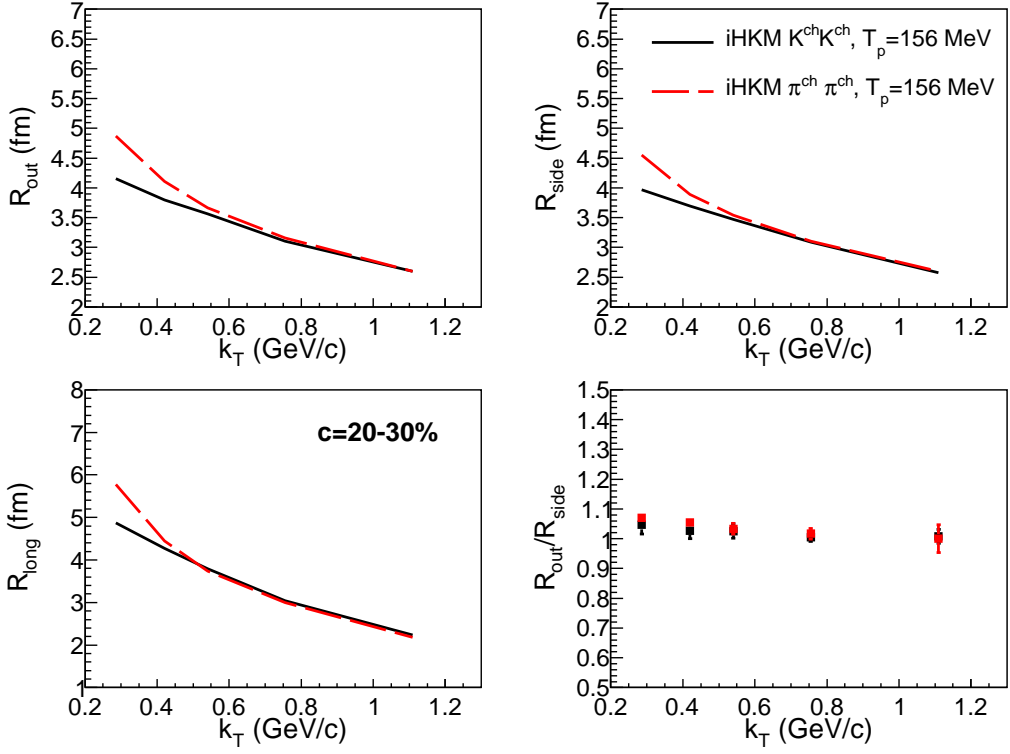}
\caption{The same as in Fig.~\ref{rad05} for $c=20-30\%$.
\label{rad2030}} 
\end{figure}

\begin{figure}
\centering
\includegraphics[bb=0 0 567 411, width=0.85\textwidth]{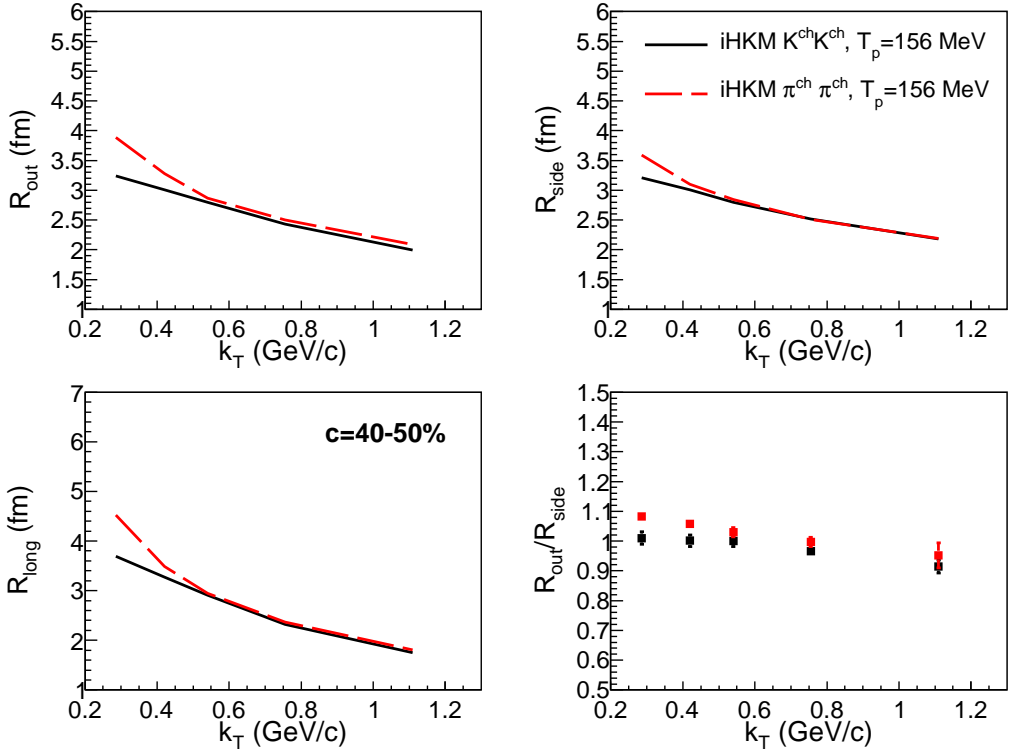}
\caption{The same as in Fig.~\ref{rad05} for $c=40-50\%$.
\label{rad4050}} 
\end{figure}

In the current study we follow the same algorithm as described in~\cite{lifetime} and perform fitting of 
the $p_T$ spectra
and \textit{long} femtoscopy radii obtained from the iHKM realistic simulations of the relativistic
heavy-ion collisions. The model consists of several modules, each describing one of the collision stages
(initial state of the system right after the nuclei have passed through each other, pre-thermal expansion 
of far-from-equilibrium system, hydrodynamical expansion of nearly thermalized matter, 
particlization and hadron cascade stage --- see~\cite{ihkm1,ihkm2} for details).
In~\cite{lhc502-ihkm} the model was tuned for the simulation of Pb+Pb collisions at the LHC energy $5.02 A$~TeV
and showed good agreement with the experimental results on different particle production observables for this energy.
The predictions for interferometry radii in the three centrality classes ($c=0-5\%$, $c=20-30\%$, and $c=40-50\%$)
were also made (see Figs.~\ref{rad05}, \ref{rad2030} and \ref{rad4050}). Thus, we have all the necessary data 
to apply our method and try to extract the maximal emission times.

\begin{figure}
\centering
\includegraphics[bb=0 0 567 411, width=0.8\textwidth]{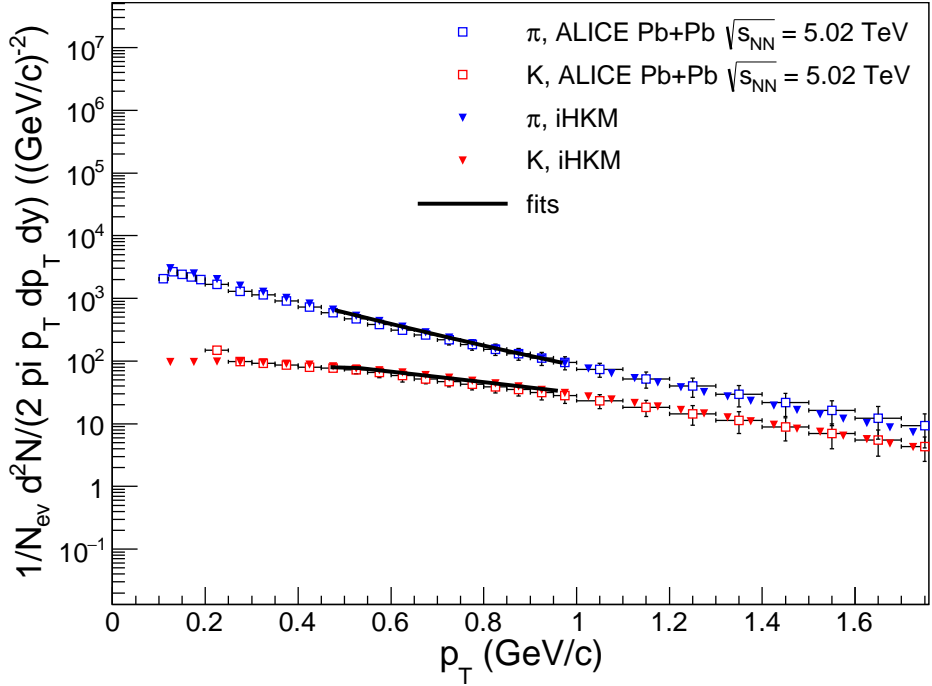}
\caption{The iHKM results for pion and kaon $p_T$ spectra compared to the ALICE data~\cite{alice-spec}
for the LHC Pb+Pb collisions at $\sqrt{s_{NN}}=5.02$~TeV ($c=0-5\%$)
together with the lines, representing a combined fit to the iHKM spectra using (\ref{specfit})
with the same effective temperature $T$ for pions and kaons.
\label{specf}} 
\end{figure}

In Fig.~\ref{specf} one can see the plot demonstrating pion and kaon $p_T$ spectra for the most central
collisions ($c=0-5\%$) calculated in iHKM together with the ALICE Collaboration experimental data~\cite{alice-spec} 
and fitting curves according to Eq.~(\ref{specfit}). The fitting is carried out in the momentum 
range $0.45<p_T<1.0$~GeV/$c$. The temperature $T$ is fixed to be a common parameter for both pion and kaon spectrum,
while the $\alpha$ and normalizing constant parameters are supposed to be different for pions and kaons.
It is worth noting that as compared to the case of $2.76 A$~TeV collisions, the fitting results are not
so stable and fluctuate depending on the used $p_T$ range and the initial constraints put on the parameters.
This can mean that the shape of the spectra for $5.02 A$~TeV Pb+Pb collisions is not so close to the
exponential one as for the lower LHC energy. In such a situation we had to compare different fitting results
and choose one of them. We finally settled on the result with the common temperature $T=138$~MeV, 
$\alpha_{\pi}=4.8 \pm 1.1$ and $\alpha_{K}=2.4 \pm 0.6$, based on several considerations, such as
that the effective temperature should not differ much from the value $T=144$~MeV,
obtained in the case of $\sqrt{s_{NN}}=2.76$~TeV, where the fit was more stable, and also that the parameter
errors and the fit's $\chi^2$ should be as small as possible.

\begin{figure}
\centering
\includegraphics[bb=0 0 567 411, width=0.8\textwidth]{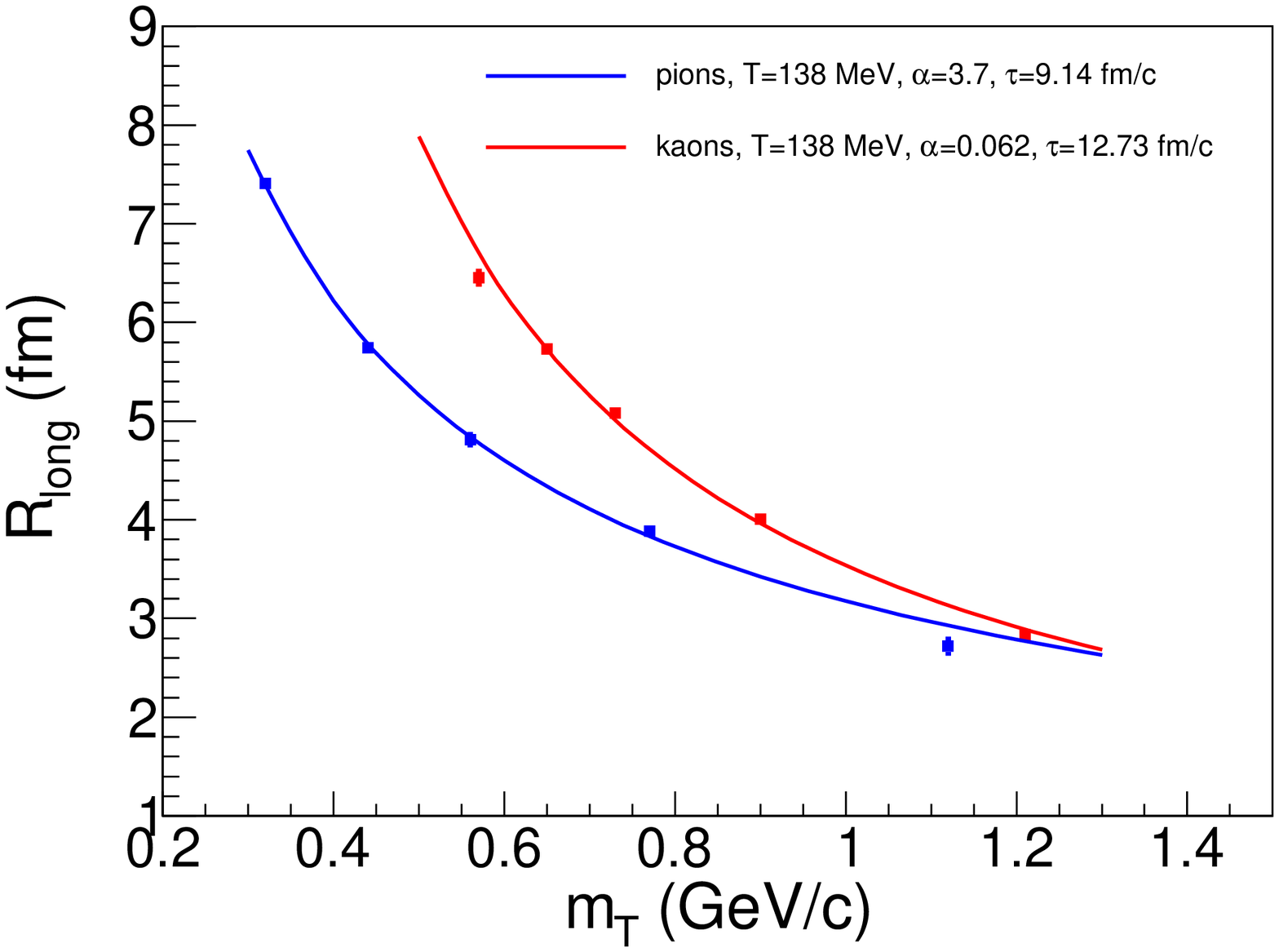}
\caption{The fitting of pion (blue squares) and kaon (red squares) femtoscopy radii,
calculated in iHKM for $c=0-5\%$ with the lines corresponding to formula (\ref{radfit}).
The common effective temperature $T=138$~MeV and the value $\alpha_{\pi}=3.7$ are taken
from the combined pion and kaon $p_T$ spectra fit with Eq. (\ref{specfit}).
The $\alpha$ value for kaons as well as the maximal emission times $\tau_\pi$ and $\tau_K$ are free parameters.
Their values extracted from the best fit are: $\alpha_{K}=0.062$, $\tau_{\pi}=9.14$~fm/$c$ 
and $\tau_{K}=12.73$~fm/$c$.
\label{rlfit05}} 
\end{figure}

Fixing then the $T$ and $\alpha$ parameters at the values found during the spectra fitting, we performed
the fitting of \textit{long} radii dependency on the pair $m_T$ (see Fig.~\ref{rlfit05}).
Similarly to \cite{lifetime} we readily obtained a good fit for pion radii and extracted the corresponding
pion maximal emission time $\tau_{\pi}=9.14$~fm/$c$, while for kaons we had to make $\alpha$ parameter
free again to get satisfactory description of the iHKM points. As a result, we obtained $\alpha_{K}=0.062$
and the maximal emission time $\tau_{K}=12.73$~fm/$c$. One can also see that the $m_T$ scaling between
pion and kaon radii is violated, as well as in $\sqrt{s_{NN}}=2.76$~TeV collisions, apparently
due to the presence of strong transverse flow and intensive hadron-hadron interactions
at the afterburner stage of the collision, as it was advocated in~\cite{lifetime}. At the same time, $k_T$ scaling 
takes place starting from $k_T \approx 0.5$~GeV/$c$ for all femtoscopy scales. Such a scaling for pion and kaon radii 
at the LHC was predicted in Ref.~\cite{pbm} and confirmed by the ALICE Collaboration~\cite{alice-mt}.  

A slight overestimation of pion radius at the highest considered $m_T$ value, $m_T=1.12$~GeV,
in fitting curve can be possibly connected with the fact that we use the approximation 
of $\tau=\tau_{m.e.}=const$ at hadronization hypersurface, applicable for soft particles with not very
high $p_T$. However quanta with $m_T$ close to 1 GeV are emitted from the side part
of the overall hadronization hypersurface, where $\tau$ values are smaller than $\tau_{m.e.}$ for soft particles.
That is why, since fitted radii values are proportional to $\tau$ (see (\ref{radfit})),
the fitting curve goes somewhat higher than the iHKM point for $m_T=1.12$~GeV.

The reason for redefining the $\alpha$ parameter for kaons for the radii fitting is non-Gaussian shape
of the correlation function. The formula (\ref{radfit}) was derived under the assumption
of small $q_{\mathrm{long}}$, and thus it describes well only the radii corresponding to
the peak part of the non-Gaussian correlation function. The $R_{\mathrm{long}}(m_T)$ fit 
with $\alpha_K$ fixed to the value extracted from the combined spectra fitting would describe the iHKM
points well if the latter were obtained from the Gaussian fits to the model correlation functions
in a narrow range for $q$, as it was demonstrated in~\cite{lifetime} (see Figs.~5, 6 from \cite{lifetime}
and the related text) for the radii extracted using the $q$ range $|q|<0.04$~GeV/$c$.
And to describe the femtoscopy radii obtained from the correlation function fits in a wider $q$
interval (we used the range $|q|<0.2$~GeV/$c$ to extract the radii presented here) one should
use a separate $\alpha_K$ value, different from that ensuring the kaon spectrum description.


\begin{figure}
\centering
\includegraphics[bb=0 0 567 411, width=0.8\textwidth]{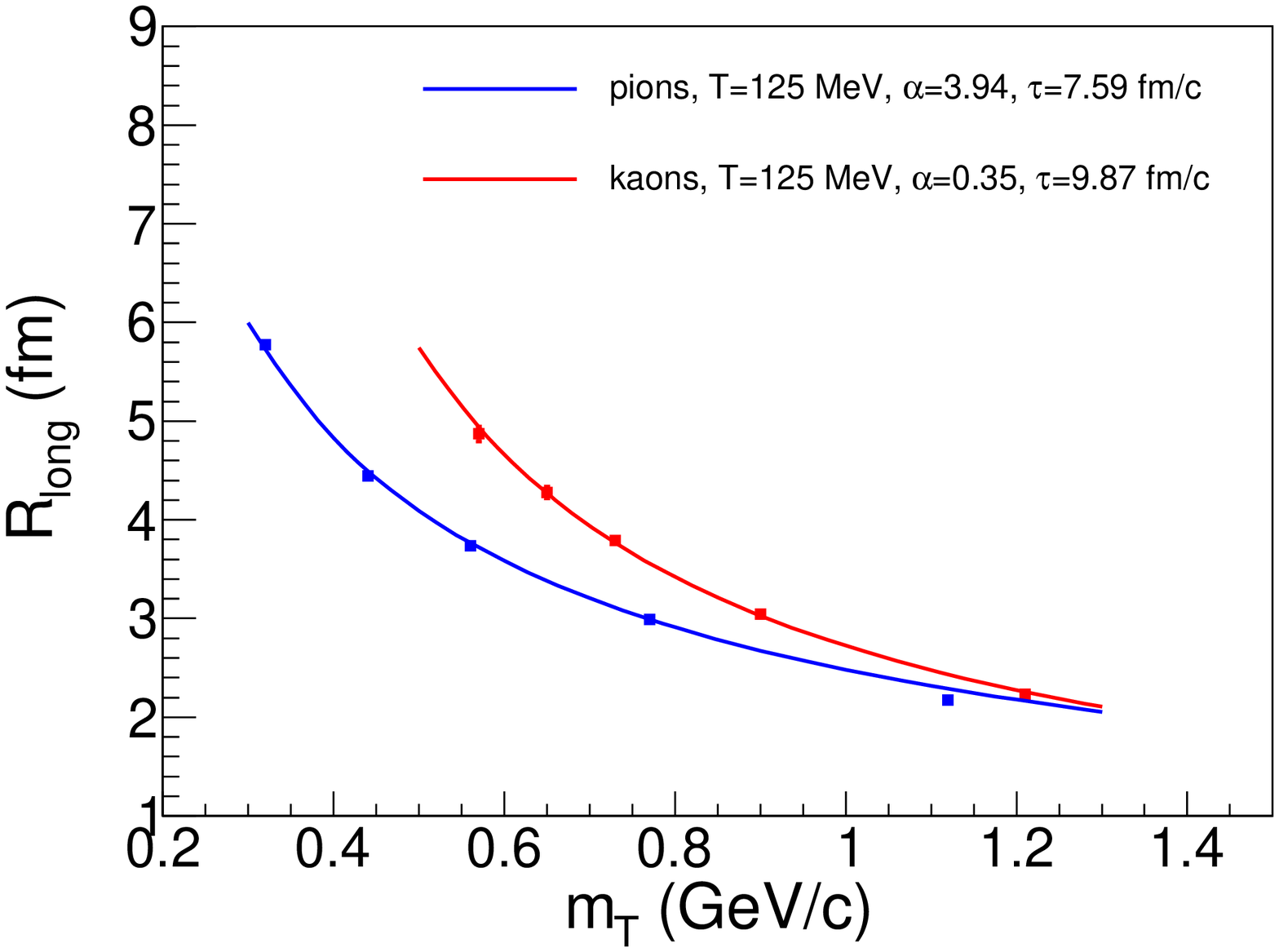}
\caption{The fitting of pion (blue squares) and kaon (red squares) femtoscopy radii,
calculated in iHKM for $c=20-30\%$ with the lines corresponding to formula (\ref{radfit}).
The common effective temperature $T=125$~MeV and the value $\alpha_{\pi}=3.94$ are taken
from the combined pion and kaon $p_T$ spectra fit with Eq. (\ref{specfit}).
The $\alpha$ value for kaons as well as the maximal emission times $\tau_\pi$ and $\tau_K$ are free parameters.
Their values extracted from the best fit are: $\alpha_{K}=0.35$, $\tau_{\pi}=7.59$~fm/$c$ 
and $\tau_{K}=9.87$~fm/$c$.
\label{rlfit2030}} 
\end{figure}

\begin{figure}
\centering
\includegraphics[bb=0 0 567 411, width=0.8\textwidth]{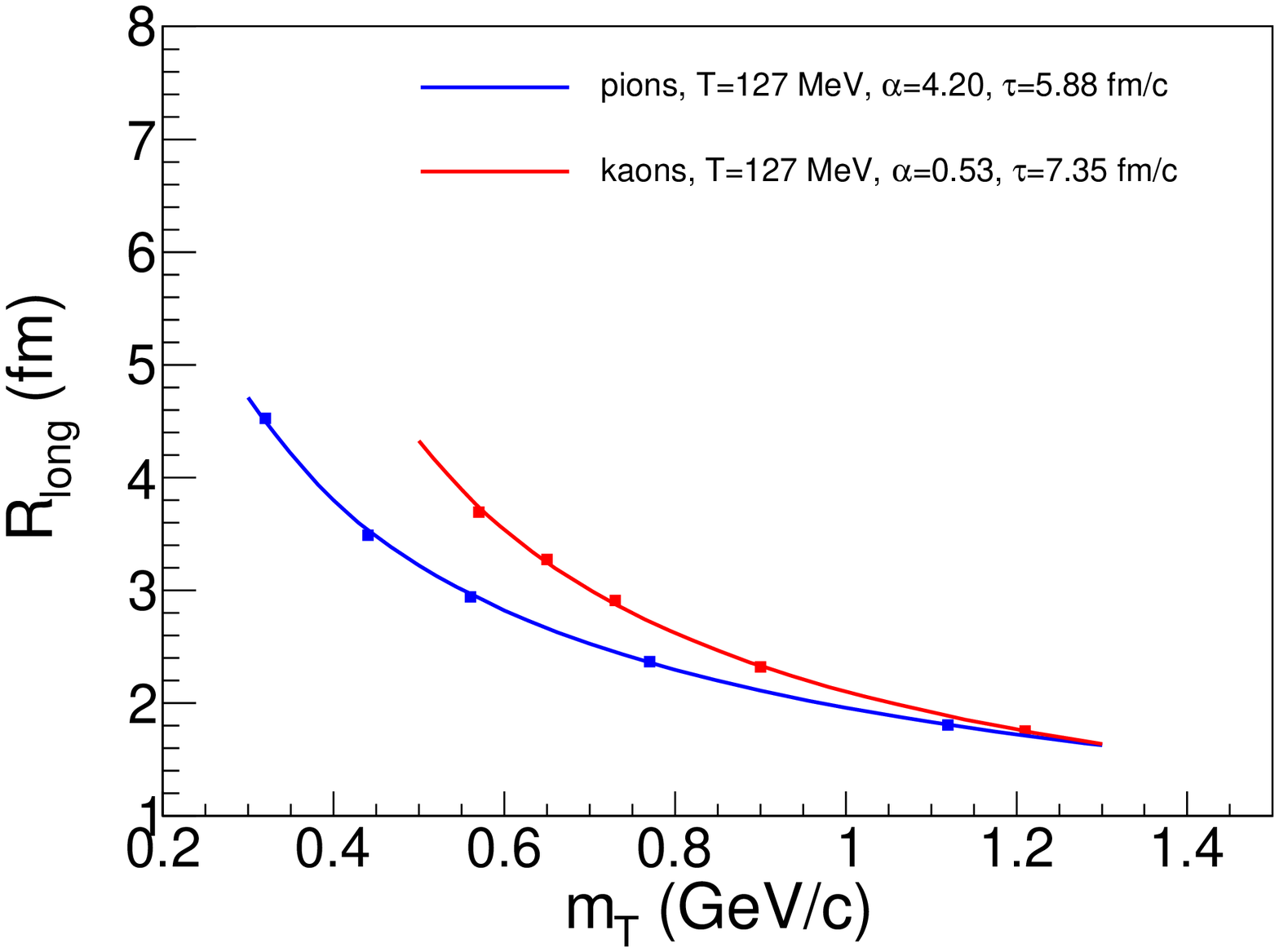}
\caption{The fitting of pion (blue squares) and kaon (red squares) femtoscopy radii,
calculated in iHKM for $c=40-50\%$ with the lines corresponding to formula (\ref{radfit}).
The common effective temperature $T=127$~MeV and the value $\alpha_{\pi}=4.20$ are taken
from the combined pion and kaon $p_T$ spectra fit with Eq. (\ref{specfit}).
The $\alpha$ value for kaons as well as the maximal emission times $\tau_\pi$ and $\tau_K$ are free parameters.
Their values extracted from the best fit are: $\alpha_{K}=0.53$, $\tau_{\pi}=5.88$~fm/$c$ 
and $\tau_{K}=7.35$~fm/$c$.
\label{rlfit4050}} 
\end{figure}

We also present the results on pion and kaon maximal emission times for the two other centrality
classes, $c=20-30\%$ and $c=40-50\%$, for which the iHKM predictions on femtoscopy radii were
made in~\cite{lhc502-ihkm}. The corresponding plots are shown in Figs.~\ref{rlfit2030} and \ref{rlfit4050}.
For the $c=20-30\%$ case the temperature parameter extracted from the combined pion and kaon
$p_T$ spectra fit is $T=125$~MeV and the corresponding $\alpha_{\pi}=3.94$.
The $\alpha_{K}=0.35$ value is again extracted from the radii fit, and the maximal emission times 
are $\tau_{\pi}=7.59$~fm/$c$ and $\tau_{K}=9.87$~fm/$c$.
Finally, in the $c=40-50\%$ case the values $T=127$~MeV and $\alpha_{\pi}=4.20$ were found from the spectra fit,
and the \textit{long} radii fit gave $\alpha_{K}=0.53$, together with the times $\tau_{\pi}=5.88$~fm/$c$ 
and $\tau_{K}=7.35$~fm/$c$.

One can note that the obtained effective temperature in case of non-central events is lower than
that for the most central collisions, as well as the corresponding maximal emission times.
The values of $\alpha$, on the contrary, are higher for the non-central collisions.
Such interrelation between the fit parameter values reflects the actual physical difference
between the systems, formed in central and non-central collisions, namely that those created in
non-central collisions live shorter, cool-down faster and develop less intensive collective flows 
during their evolution.

\begin{figure}
\includegraphics[width=0.9\textwidth]{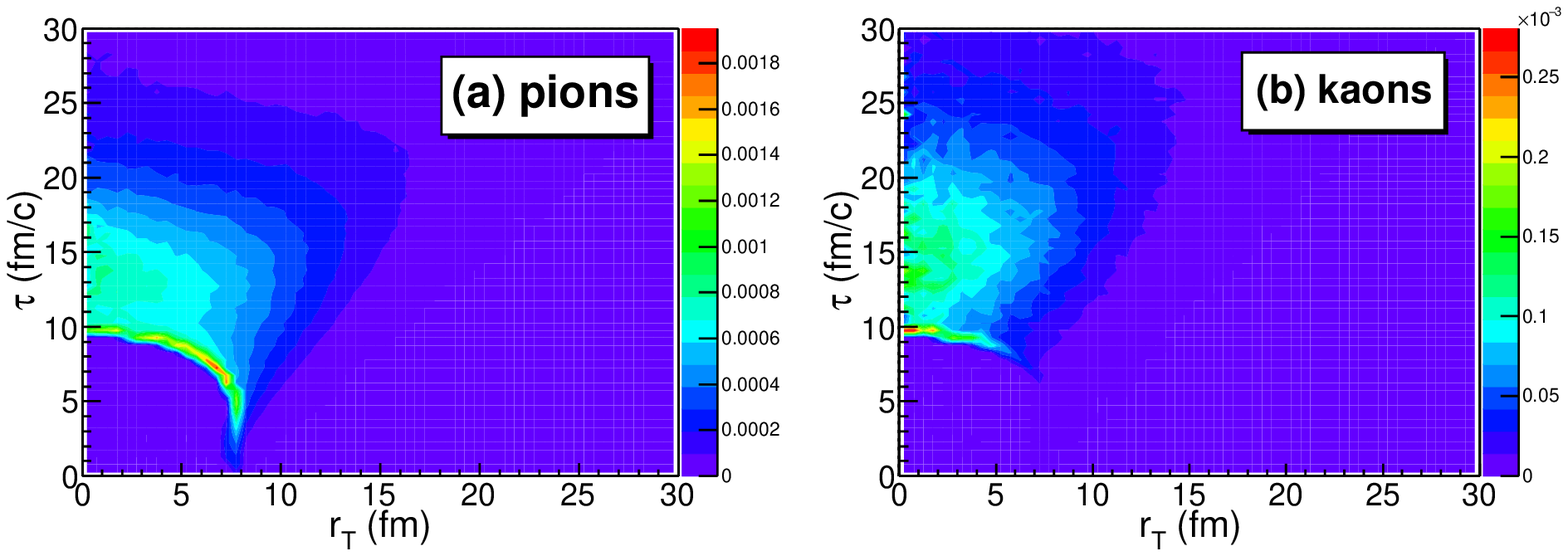}
\caption{The emission functions per units of
space-time and momentum rapidities averaged over momentum angles $ g(\tau,
r_T,p_T)$ [fm$^{-3}$] for pions~(a) and kaons~(b) obtained from
the iHKM simulations of the LHC Pb+Pb collisions at the energy
$\sqrt{s_{NN}}=5.02$~TeV, $0.2<p_T<0.3$~GeV/$c$, $|y|<0.5$,
$c=0-5$\%. } 
\label{emiss1}
\end{figure}

\begin{figure}
\includegraphics[width=0.9\textwidth]{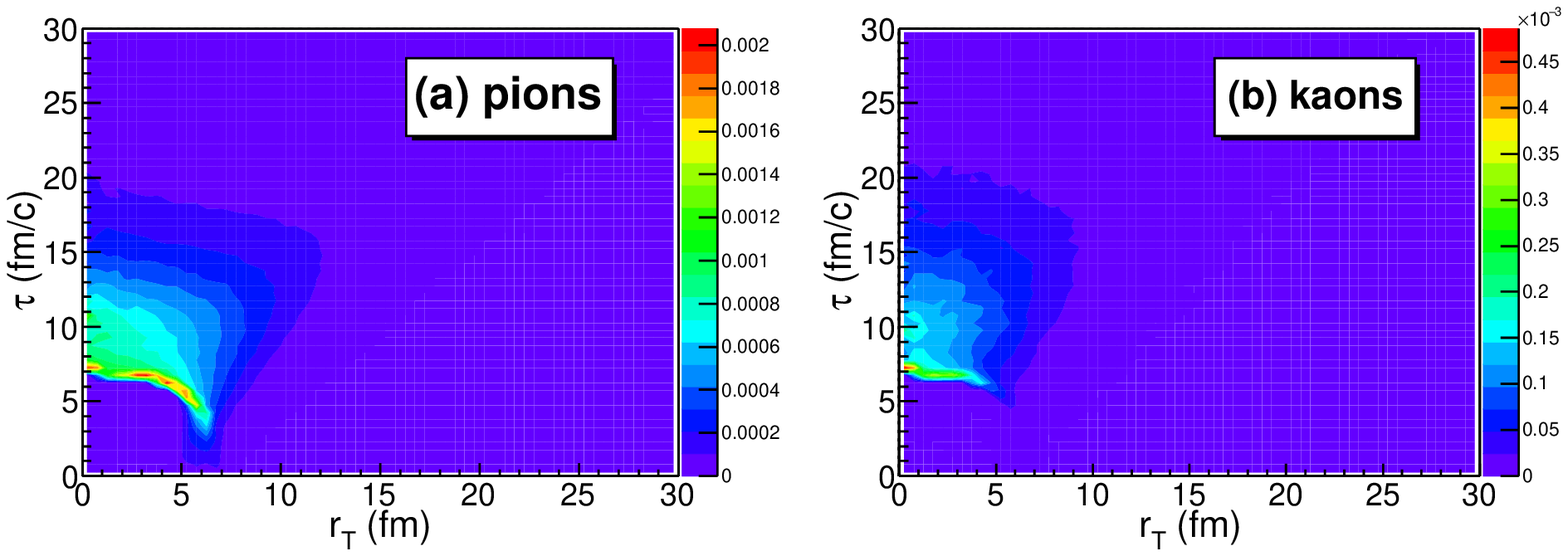}
\caption{The same as in Fig.~\ref{emiss1} for the events from $c=20-30$\% centrality class. } 
\label{emiss2}
\end{figure}

\begin{figure}
\includegraphics[width=0.9\textwidth]{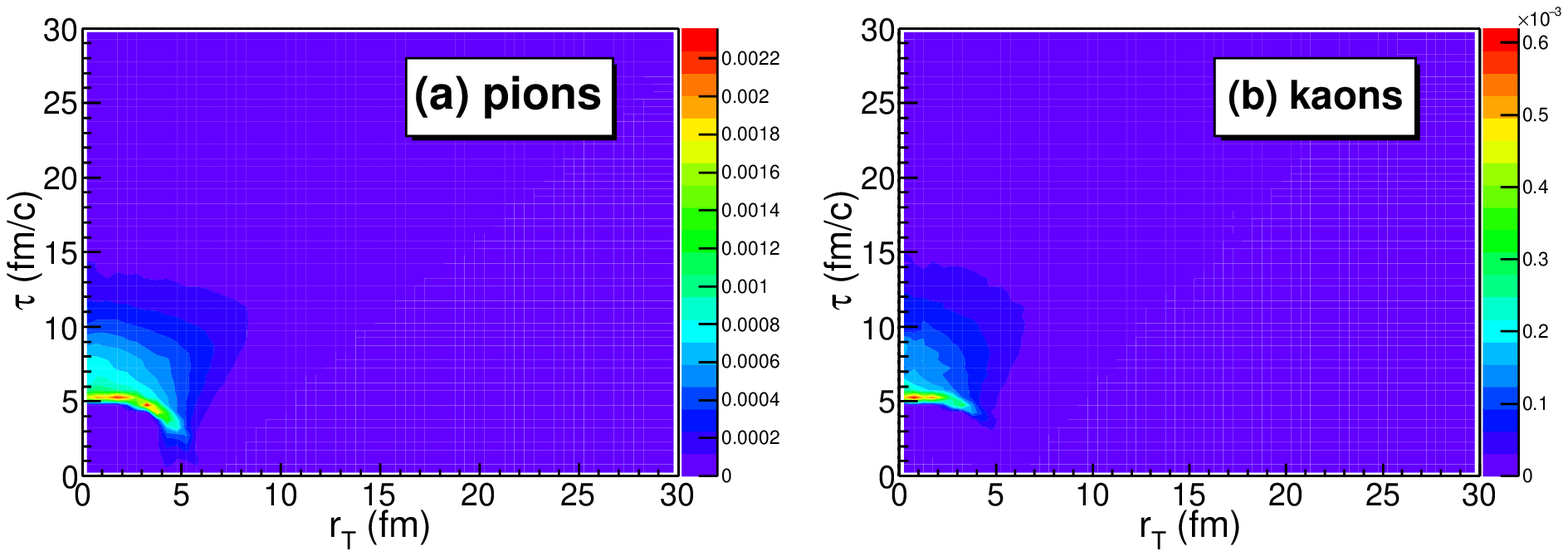}
\caption{The same as in Fig.~\ref{emiss1} for the events from $c=40-50$\% centrality class.  } 
\label{emiss3}
\end{figure}

Additionally, in Figs.~\ref{emiss1}-\ref{emiss3} we demonstrate the plots for the averaged emission functions
of pions and kaons for the three considered centrality classes, which allows to qualitatively identify the regions 
of the maximal emission for particles of each species (with $0.2<p_T<0.3$~GeV/$c$) and in such a way 
to approximately estimate the corresponding effective maximal emission times~$\tau$. 

One can see, that the maximal emission time values, previously obtained from the fits, e.g. for $c=0-5\%$ events,
$\tau_{\pi}=9.14$~fm/$c$ and $\tau_{K}=12.73$~fm/$c$, are in agreement 
with the presented emission pictures, since according to the plot, the maximum of pion emission
should be close to the particlization time in the center of the system, $\tau \approx 10$~fm/$c$,
and for kaons the effective $\tau_K$ value should be between the two emission maxima, seen on the plot 
(more clearly in numerical representation),
namely between $\tau \approx 10$~fm/$c$ and $\tau \approx 15$~fm/$c$. The second in time maximum is conditioned by 
$K^{*}(892)\rightarrow \pi + K$ decays, as it was earlier explained in Ref.~\cite{lifetime} 
and in fact was confirmed by the results of the ALICE Collaboration~\cite{alice-mt}. 

Quite similar situation takes place also for non-central collisions (see Figs.~\ref{emiss2}, \ref{emiss3}): the
times of maximal emission for pions, $\tau_{\pi}=7.59$~fm/$c$ and $\tau_{\pi}=5.88$~fm/$c$, extracted from the fits
are close to the central parts' particlization times, $\tau \approx 7.5$~fm/$c$ and $\tau \approx 5$~fm/$c$, 
following from the emission pictures. For kaons the obtained times of maximal emission, $\tau_{K}=9.87$~fm/$c$ 
and $\tau_{K}=7.35$~fm/$c$, are, similarly to the central collision case, higher than those of pions by 
about $2-3$~fm/$c$, i.e. are between the particlization time and the time of $K^{*}$ resonance decay, whose
lifetime is about $4-5$~fm/$c$ and which forms a second kaon emission maximum on the radiation plots (however,
for non-central collisions, especially for the $c=40-50$\% case, this second maximum is less pronounced
than in central events, maybe due to smaller number of produced particles in non-central collisions).

\section{Conclusions}

The times of the maximal emission of kaons and pions in the Pb+Pb collisions at the LHC energy $5.02 A$~TeV
were estimated based on the transverse momentum spectra and \textit{long} femtoscopy radii fitting
with the analytical formulas accounting for the presence of transverse collective flow.
Comparing the fitting results for different centrality classes, one can observe that the maximal emission times,
effective temperatures, and flow intensities are smaller in non-central events, than in central ones.
The fitting results are in a qualitative agreement with the regions of the most intensive particle
emission seen on the averaged emission function plots.
The stability of the spectra fits is worse as compared to the case of the lower LHC energy $2.76 A$~TeV,
which may be due to noticeable deviations of the spectrum shape from the exponential.

In this note, we found that, similarly to LHC Pb+Pb collisions at the energy $2.76 A$~TeV, kaons radiate later 
than pions at all centralities also at the energy $5.02 A$~TeV. The reason, again, is mostly decays of $K^{*}$ 
resonance. The intensive hadron-hadron scatterings at the afterburner stage of the collision along with very intensive 
transverse flow results in breaking of $m_T$ scaling between pion and kaon interferometry radii.  
At the same time, we again, as for the smaller LHC energy, predict $k_T$ scaling at not very small $k_T$ for pion and 
kaon femto-scales also for Pb+Pb collisions at currently the highest energy. 
We are looking forward to the corresponding 
results of femtoscopic analysis from the LHC Collaborations for the energy $5.02 A$~TeV.

\begin{acknowledgments}
The authors express their sincere gratitude to L.~Malinina and G.~Romanenko for their interest in this work and useful discussions. The research was carried out within the NAS of Ukraine Targeted research program ``Collaboration in advanced 
international projects on high-energy physics and nuclear physics'', 2021, 
Agreement \textnumero 7-2021 between the NAS of Ukraine and BITP of NASU. 
\end{acknowledgments}


\begin{thebibliography}{99}

\bibitem{Lisa} 
M. A. Lisa, S. Pratt, R. Soltz, and U. Wiedemann, Ann. Rev. Nucl. Part. Sci. {\bf  55}, 357 (2005).

\bibitem{hlength1} 
Yu. M. Sinyukov, Nucl. Phys. A {\bf 566}, 589 (1994);\\
Yu.M. Sinyukov, in \textit{Hot Hadronic Matter: Theory and Experiment},
edited by J. Letessier, H. H. Gutbrod and J. Rafelski (Plenum, New York, 1995), p. 309.

\bibitem{hlength2} 
S.V. Akkelin, Yu.M. Sinyukov, Phys. Lett. B {\bf 356}, 525 (1995).

\bibitem{MakSin}        
A.N. Makhlin and Yu.M. Sinyukov, Sov. J. Nucl. Phys. {\bf  46}, 354 (1987); \\
Yu.M. Sinyukov, Nucl. Phys. A {\bf 498}, 151c (1989).

\bibitem{hbt-puzzle1}
M.S. Borysova, Yu.M. Sinyukov, S.V. Akkelin, B. Erazmus, Iu.A. Karpenko, Phys. Rev. C \textbf{73}, 024903 (2006).

\bibitem{hbt-puzzle2}
Iu. A. Karpenko and Yu. M. Sinyukov, Phys. Lett. B \textbf{688}, 50 (2010).

\bibitem{lifetime}
Yu.M. Sinyukov, V.M. Shapoval, V.Yu. Naboka, Nucl. Phys. A \textbf{946}, 227 (2016).

\bibitem{sourcefunc} 
V.~M.~Shapoval, Yu.~M.~Sinyukov, Iu.~A.~Karpenko, Phys. Rev. C \textbf{88}, 064904 (2013).

\bibitem{lednicky}
R. Lednicky, V. L. Lyuboshits, B. Erazmus, and D. Nouais, Phys. Lett. B \textbf{373}, 30–34 (1996).

\bibitem{kiesel}
A. Kisiel, Phys. Rev. C \textbf{81}, 064906 (2010).

\bibitem{HKM} 
Yu.M. Sinyukov, S.V. Akkelin, and Y. Hama, Phys. Rev. Lett. \textbf{89}, 052301 (2002).

\bibitem{HKM1} 
S.V. Akkelin, Y. Hama, Iu.A. Karpenko, Yu.M. Sinyukov, Phys. Rev. C \textbf{78}, 034906 (2008).

\bibitem{Oslo} Y.V. Kravchenko \textit{et al.}, Phys. Scr. {\bf 96}, 104002 (2021).

\bibitem{alice-mt}
J. Adam \textit{et al.} (ALICE Collaboration), Phys. Rev. C \textbf{96}, 064613 (2017).

\bibitem{ihkm1}
V.Yu. Naboka, S.V. Akkelin, Iu.A. Karpenko, Yu.M. Sinyukov, Phys. Rev. C \textbf{91}, 014906 (2015).

\bibitem{ihkm2} 
V.Yu. Naboka, Iu.A. Karpenko, Yu.M. Sinyukov, Phys. Rev. C \textbf{93}, 024902 (2016).

\bibitem{rhic-ihkm}
M.D. Adzhymambetov, V.M. Shapoval, Yu.M. Sinyukov, Nucl. Phys. A \textbf{987}, 321 (2019).

\bibitem{lhc502-ihkm}
V.M. Shapoval, Yu.M. Sinyukov, Phys. Rev. C \textbf{100}, 044905 (2019).

\bibitem{spec-form1}
S.V. Akkelin, Y. Hama, Iu.A. Karpenko, Yu.M. Sinyukov, Phys. Rev. C \textbf{78}, 034906 (2008).

\bibitem{spec-form2}
Yu.M. Sinyukov, S.V. Akkelin, Y. Hama, Iu.A. Karpenko, Act. Phys. Pol. B \textbf{40}, 1025 (2009).

\bibitem{tau-const1}
S.V. Akkelin, Yu.M. Sinyukov, Phys. Lett. B \textbf{356}, 525 (1995).

\bibitem{tau-const2}
S.V. Akkelin, Yu.M. Sinyukov, Z. Phys. C \textbf{72}, 501 (1996).
 
\bibitem{Tolstykh} 
Yu.M. Sinyukov, S.V. Akkelin, A.Yu. Tolstykh, Nucl. Phys. A  \textbf{72} 278c (1996).

\bibitem{alice-spec} 
N. Jacazio (for the ALICE Collaboration), Nucl. Phys. A \textbf{967}, 421 (2017).

\bibitem{pbm} 
V.M. Shapoval, P. Braun-Munzinger, Iu.A. Karpenko, Yu.M. Sinyukov, Nucl. Phys. A \textbf{929}, 1 (2014).  

\end{thebibliography}
\end{document}